\documentclass[aps,pre,reprint,superscriptaddress,letterpaper,twocolumn]{revtex4-1}

\usepackage{color}
\usepackage{amssymb}
\usepackage{amsmath,mathrsfs}
\usepackage{braket}
\usepackage{bm} 

\usepackage[ruled]{algorithm2e}
\usepackage{algorithmic}
\usepackage{setspace,soul}
\usepackage{graphicx}

\usepackage{grffile}
\usepackage{subfigure}

\usepackage{hyperref}

\usepackage{cleveref}
\usepackage[utf8]{inputenc}
\usepackage{geometry}
\usepackage{float}

\usepackage{xparse}
\ExplSyntaxOn
\NewDocumentCommand{\mref}{m}{\quinn_mref:n {#1}}
\seq_new:N \l_quinn_mref_seq
\cs_new:Npn \quinn_mref:n #1
 {
  \seq_set_split:Nnn \l_quinn_mref_seq { , } { #1 }
  \seq_pop_right:NN \l_quinn_mref_seq \l_tmpa_tl
  ( 
  \seq_map_inline:Nn \l_quinn_mref_seq
    { \ref{##1},\nobreakspace } 
  \exp_args:NV \ref \l_tmpa_tl 
  ) 
 }
\ExplSyntaxOff

\begin{document}

\title{Gaussian Closure Scheme in the  Quasi--Linkage Equilibrium Regime of Evolving Genome Populations}

\author{Eugenio Mauri}
\email{emauri@clipper.ens.psl.eu}
\affiliation{Laboratory of Physics of the Ecole Normale Sup\'{e}rieure, CNRS UMR 8023 and PSL Research, 24 rue Lhomond, 75231 Paris cedex 05, France}

\author{Simona Cocco}
\email{simona.cocco@phys.ens.fr}
\affiliation{Laboratory of Physics of the Ecole Normale Sup\'{e}rieure, CNRS UMR 8023 and PSL Research, 24 rue Lhomond, 75231 Paris cedex 05, France}

\author{R\'emi Monasson}
\email{monasson@lpt.ens.fr}
\affiliation{Laboratory of Physics of the Ecole Normale Sup\'{e}rieure, CNRS UMR 8023 and PSL Research, 24 rue Lhomond, 75231 Paris cedex 05, France}

\begin{abstract}
Describing the evolution of a population of genomes evolving in a complex fitness landscape is generally very hard. We here introduce an approximate Gaussian closure scheme to characterize analytically the statistics of a genomic population in the so-called Quasi--Linkage Equilibrium (QLE) regime, applicable to  generic values of the rates of mutation or recombination and fitness functions.  The Gaussian approximation is illustrated on a short-range fitness landscape with two far away and competing maxima. It unveils the existence of a phase transition from a broad to a polarized distribution of genomes as the strength of epistatic couplings is increased, characterized by slow coarsening dynamics of competing allele domains. Results of the closure scheme are corroborated by numerical simulations.
\end{abstract}

\date{\today}

\maketitle

\section{Introduction}

Understanding how genomes stochastically change as a result of selection, mutation and sexual recombination processes is the central goal of evolutionary biology. While the case of selection acting independently on the different alleles, a regime called linkage equilibrium (LE), has been intensively studied describing the dynamics of a population of genomes evolving in a complex fitness landscape, characterized by multiple and strong couplings between alleles, remains  a formidable challenge to population geneticists and statistical physicists~\cite{Barton997}. An important step beyond LE was done by Kimura in '65 \cite{Kimura65} for a 2-loci system at high recombination rate $r$, a regime called quasi-linkage equilibrium (QLE). Recently, Neher and Shraimann extensively studied the same regime for multi-loci systems \cite{Neher_QLE}. In QLE the distribution of genomes in a population may reach a stationary regime in which the pairwise correlations between alleles are weak, proportional to the fitness coupling between these two alleles and to the inverse $1/r$ of the recombination rate. To what extent this regime remains present or valid at low recombination remains unclear. 

The irruption of massive sequence data, following the rapid drop in sequencing cost, has brought additional interest to such questions. It is now legitimate to ask how relevant parameters (such as mutation or recombination rates, or the full fitness function) can be extracted from sequences. An example can be found in \cite{HIV_rec}, where the effective recombination rate of HIV was inferred based on a specific model for intra-patient evolution. The large--$r$ relation between correlations and epistatic coupling was also used to infer fitness contribution from data in \cite{Gao_2019}. However, incorporating evolutionary models into inverse statistical approaches, such as the ones developed for extracting structural and functional constraints from protein sequences \cite{coccorev} remains a largely open issue so far.

The daunting complexity of the relationship between fitness determinants and  the statistical properties of genomic data calls for the design of accurate and practical approximation schemes. We hereafter consider a Gaussian closure scheme for the hierarchy of moments of the allele distributions. 
Various other proposals of general, approximate frameworks in the context of the dynamics of populations can be found in literature, e.g. \cite{BartonTurelli94,Buerger2000} and, more recently, \cite{Jain2017,Barghi2020}. In particular Barton and Turelli introduced general recursion equations for multi-locus cumulants, which they apply to the case of interchangeable additive traits under strong selection \cite{BartonTurelli94}. While our approximation is less general than Barton and Turelli's as it relies on the first and second moments only, it is a practical tool to deal with cases in which traits are neither interchangeable, {\em i.e} loci contribute differently to fitness, nor additive, as found in predominantly epistatic fitness landscapes.

As we will show, our Gaussian scheme allows us to bypass the large--$r$ expansion made in \cite{Kimura65,Neher_QLE}, and study the existence of QLE in a larger range of recombination and mutation rates. We apply our approximation to a specific fitness model in which two very different genomes have maximal and equal fitness values, and with short-range (along the genome) and strong epistatic contributions. As the mutation and recombination rates are varied a phase transition takes place within the QLE regime, which separates a disordered (paramagnetic) regime in which alleles are not polarized along any of the two genomes and a polarized (ferromagnetic) phase in which, after coarsening of alternating allele domains, one of the two genomes dominates in the population. 

\section{Model for the stochastic evolution of a genome population}
We consider a population of $N$ individuals, whose genomes $\textbf{s}^{(k)}=\{s_0^{(k)}, s^{(k)}_1, ..., s^{(k)}_{L-1}\}$, with $k=1,...N$, include $L$ loci.  Each locus $i=0,...,L-1$ can carry one out of two states (alleles), $s_i=\pm 1$. Genomes stochastically evolve over time $t$, due to selection, mutation and sexual recombination processes. We will denote by $\langle G\rangle (t)$ the average of any function $G({\bf s})$ over the genomes in the population at time $t$,
\begin{equation}
\braket{G} (t) = \frac 1N \sum_{k}G\big(\textbf{s}^{(k)}\big) \ .
\end{equation}

{\em Selection.} The number of individuals carrying sequence $\textbf{s}$ grow inside the population according to the growth rate $F(\textbf{s})$, called \textit{fitness}. The fitness function can be expanded in terms of the state variables,
	\begin{equation}\label{fit1}
	    F(\textbf{s}) = F_0 + \sum_i f_i\, s_i + \sum_{i<j}f_{ij}\,s_i\,s_j\ ,
	\end{equation}
where we neglect all the interactions of order equal to, or higher than three, and consider  only local ($f_i$) and pairwise epistatic ($f_{ij}$) contributions. 
In practice, selection can be enforced in simulations as follows. Time is discretize into very small time steps $\Delta t$. At each  time step selection results in a re-sampling of the available sequences according to a Boltzmann probability distribution with the weights defined by the fitness of each sequence, i.e. $p(\textbf{s}_k)\propto  {e^{F(\textbf{s}_k)}}$ with $k = 1,\dots , N$. We make sure that the population size, $N$, is kept fixed. Note that the constant $F_0$ plays no role and can be set to zero.

{\em Mutation.} Each state $s_i$ can flip its value, $s_i\to -s_i$, with rate (probability per unit of time) $\mu$. This rate is supposed to be uniform across loci $i$ and individuals.

{\em Recombination.} For each infinitesimal time interval $\Delta t$, with probability $r\Delta t/N_{pairs}$, a pair of individuals is chosen uniformly at random  among all $N_{pairs}=\frac 12 N(N-1)$ pairs, where $r$ is the recombination rate. These two individuals, called mother \textit{(m)} and father \textit{(f)}, undergo outcrossing. Two new genotypes $\bf s$, $\bf s'$ are formed by inheriting some loci from the genome of the mother, $\textbf{s}^{(m)}$, and the complement from the father with genome $\textbf{s}^{(f)}$. This process is described by the inheritance vector $\boldsymbol\xi=\{\xi_i\}$, with $\xi_i \in \{0,1\}$: the $i$-th allele of $\bf s$ is inherited from the mother if $\xi_i = 1$  and from the father if $x_i=0$, and vice versa for $\bf s'$. Hence, $s_i = s_i^{(m)}\, \xi_i + s_i ^{(f)}\,(1-\xi_i)$. The inheritance vector $\boldsymbol\xi$ is itself stochastic, and we denote by $C({\boldsymbol\xi})$ its probability. In the following, we will need to consider in particular the recombination correlation $c_{ij}\equiv\sum_{\boldsymbol\xi}C(\boldsymbol\xi)\left[\xi_i(1-\xi_j)+(1-\xi_i)\xi_j \right]$, which represents the probability that loci $i$ and $j$ take their alleles $s_i$ and $s_j$ from different parents.

\section{Equations for the spin moments and closure scheme} 
The above dynamical processes entirely characterizes the stochastic evolution of the genome population. We will hereafter track the first two cumulants of the allele distribution over time,
\begin{equation}
\chi_i (t) = \langle s_i\rangle (t) \ , \quad \chi_{ij} (t) = \langle s_i s_ j\rangle (t) - \langle s_i\rangle (t)\langle s_j\rangle (t) \ ,
\end{equation}
where the averages are computed over the genomes in the population at time $t$. Note that $\chi_{ii}(t)=1-\chi_i(t)^2$.

In the infinite size limit ($N\to \infty$), these moments obey deterministic first-order differential equations derived by Neher and Shraimann \cite{Neher_QLE}. Unfortunately these equations are not closed, as they involve higher-order moments in the allele variables. Our closure scheme consists in replacing these high-order moments with the values they would have if the distribution of the $s_i$'s were Gaussian, {\em i.e.} in neglecting all cumulants of order $\ge3$. Qualitatively speaking, we assume a quasi-species-like scenario, in which the population is distributed around an average sequence given by $\boldsymbol\chi=\{\chi_0,...,\chi_{L-1}\}$, with some fluctuations encoded in the covariance matrix of components $\chi_{ij}$. As both the average sequence and the covariance matrix depend on $t$, the approximation offers an analytical way to track the genome population in the course of evolution. 

 Within the Gaussian closure scheme  we obtain the following set of $\frac 12 L(L+1)$ coupled equations (with the convention $f_{ij}=f_{ji}$ if $i\ne j$ and $f_{ii}=0$), see Appendix A for derivation,
   	\begin{equation}
	\dot{\chi}_i = \sum_{j} \chi_{ij} \left[ f_j + \sum_{k}f_{jk}\chi_k-2f_{ij}\chi_i\right]-2\mu\,\chi_i\, , 
	\label{eq:QLE}
	\end{equation}
and, for $i\neq j$,
	\begin{eqnarray}
		\dot{\chi}_{ij} &=&  -2\chi_{ij}\left[f_j\chi_j+f_i\chi_i\right] +  \sum_{k\ne l}  \chi_{ik}\, f_{kl}\, \chi_{jl} \nonumber\\
		&-&2\chi_{ij}\sum_{k}\big[ f_{ki} (\chi_{ik} + \chi_i\chi_k ) +f_{kj} (\chi_{jk} + \chi_j\chi_k )\big]  \nonumber\\
		&+& 2f_{ij}\chi_{ij}\left[\chi_{ij} + 2 \chi_i\chi_j\right]- (4\mu+rc_{ij})\chi_{ij}\, . \label{eq:corr}
	\end{eqnarray}

\subsection{Case of additive fitness} In the absence of any epistatic contribution to the fitness ($f_{ij}=0$ for all $i\neq j$), equation \eqref{eq:corr} for the second cumulant becomes $\dot{\chi}_{ij} = -  (2 f_j\chi_j+2 f_i\chi_i + 4\mu+rc_{ij})\chi_{ij}$, whose only fixed point is $\chi_{ij}=0$. Hence, recombination and mutations contribute to erase correlation between alleles, a scenario known as LE. The resulting value of the first moment is root of $f_i(1-\chi_i^2) - 2\mu\chi_i=0$; it smoothly interpolates between $\chi_i=2f_i/\mu$ at large mutation rate and $\chi_i=\text{sign}(f_i)$ at vanishingly small $\mu$, which corresponds to monoclonality.

\subsection{Case of random epistatic contributions: comparison to numerical simulations {and strong recombination theory of QLE}} 
\begin{figure}
	\centering
	\includegraphics[width=\linewidth]{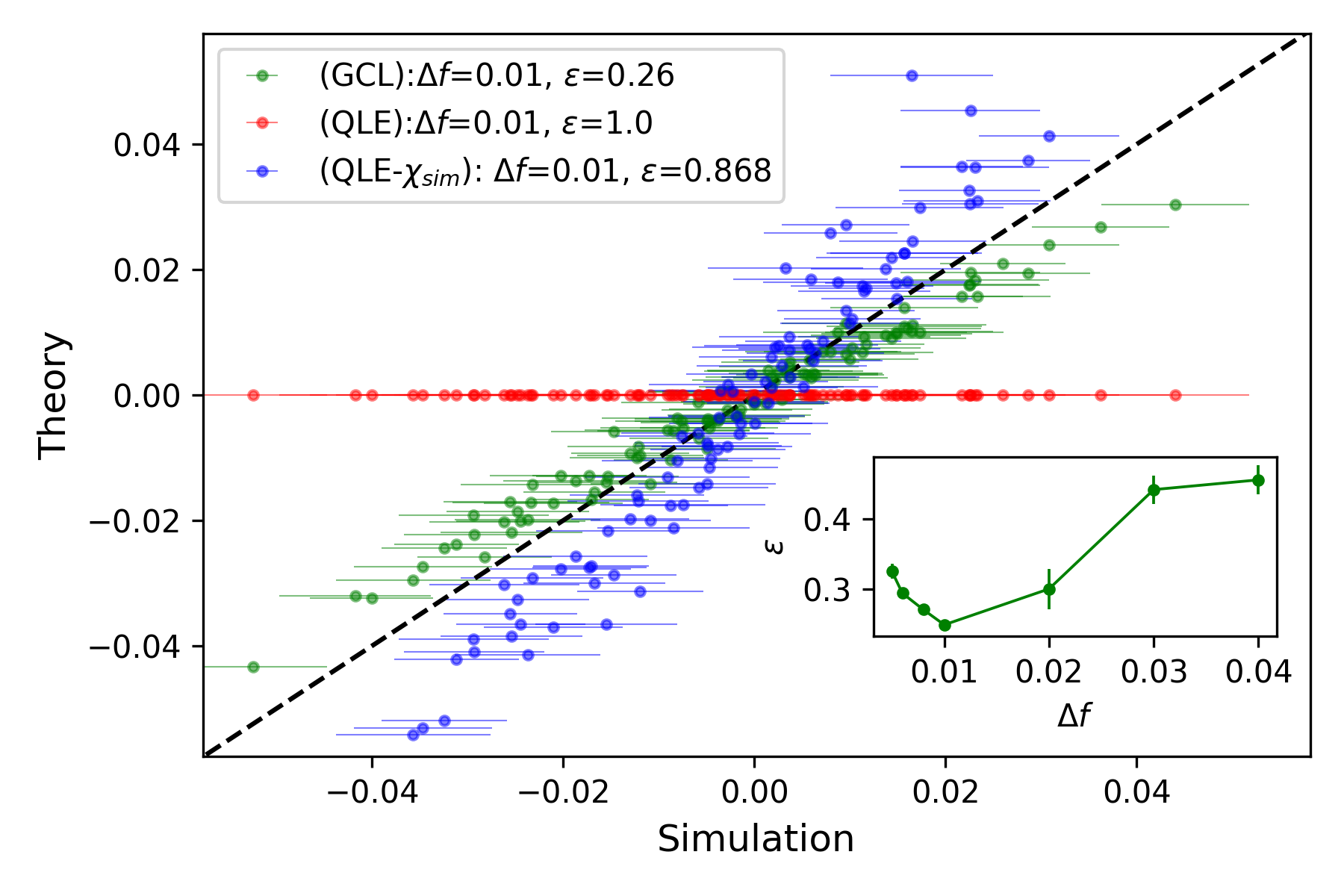}
	\caption{{Scatter plot of the correlations $\chi_{ij}$ obtained from the Gaussian closure scheme (Eqs.~\mref{eq:QLE,eq:corr}) (green) or their estimates from large--$r$ QLE theory (red: eqs.~(27) of \cite{Neher_QLE} for $\chi_i$ and $\chi_{ij}$; blue: same equations for $\chi_{ij}$ only, replacing the $\chi_i$'s with their actual values $\chi_i^{sim}$ found in simulation to account for mutations) vs.  simulations of finite populations (averaged over 500 generations once stationarity is reached) in the case of fully epistatic fitness landscape, see text and parameters in figure. Theoretical estimates of $\chi_{ij}$ are obtained by solving the dynamical system~\mref{eq:QLE,eq:corr} within Euler approximation.
			Inset: relative error $\epsilon$ (\ref{eps3}) as a function of the standard deviation $\Delta f$ of the epistatic couplings. Parameters: $L=15$, $\mu=0.08$, $r=0.85$, $\Delta h = 0.01$ and $c_{ij}=\frac 12 (1-\rho^{|i-j|})$ with $\rho = 0.6$.}}
	\label{fig:SK_theory_vs_sim}
\end{figure}

As a first, empirical test of the quantitative accuracy of the Gaussian closure scheme we have simulated the stochastic evolution of a population of $N=10,000$ individuals. {The fitness is defined by eqn \mref{fit1}, with random quenched local biases $f_i$ and epistatic couplings $f_{ij}$ drawn from a normal law with zero mean and variances equal to, respectively, $\Delta h^2$ and $\Delta f^2$}.
In Fig.~\ref{fig:SK_theory_vs_sim} we show the scatter plot of  the correlations obtained from simulations once stationarity is reached, $\chi_{ij}^{sim}$, vs. the solutions of \mref{eq:corr}, $\chi_{ij}$, for one generic sample. The Gaussian Ansatz correctly estimates small correlations, but is less accurate for large ones (in absolute values). {It performs substantially better that the QLE approximation in the high-$r$ regime from \cite{Neher_QLE}, see red and blue dots and caption of Fig.~\ref{fig:SK_theory_vs_sim} for details.} In the inset of Fig.~\ref{fig:SK_theory_vs_sim} we show the normalized mean squared error,
\begin{equation}\label{eps3}
\epsilon = \sqrt{\displaystyle{\sum_{i<j}\big|\chi_{ij}-\chi^{sim}_{ij}\big|^2}\big/{\displaystyle{\sum_{i<j}\big(\chi_{ij}^{sim}\big)^2}}}\,,
\end{equation}
as a function of $\Delta f$ {in the Gaussian closure scheme}. For these values of $\mu$ and $r$, we obtain decent estimates of the correlations {as long as $\Delta f$ is not too large, see below.}

\subsection{{Self-consistent condition for the validity of the Gaussian closure}}
{The computation of high order moments, which should vanish if the Gaussian approximation were exact, is sketched in Appendix~B. We obtain in particular that moments of order $k$ are, for large $k (\le L)$, bounded from above by $\sim \exp( \log k!! + \frac k2 \log |\chi_{ij}|+ O(\log L))$, where $|\chi_{ij}|$ denotes the order of magnitude of the 2-point correlations. Due to the asymptotic scaling of the double factorial, we conclude that $k^{th}$ moment decays exponentially with $k$  if
	\begin{equation} \label{sc}
	L\times |\chi_{ij}| < 1 \ .
	\end{equation}
	In other words, the total amount of correlations that a site, say, $i$, `receives' from the $(L-1)$ other sites, say, $j$ should be smaller than unity. This condition is sufficient and not necessary, as possible cancellation between terms of opposite signs have not been considered. It is in addition self-consistent as it must be fulfilled by the 2-point correlations computed within the Gaussian closure scheme. We will check that condition (\ref{sc}) is satisfied for the non-trivial model with epistatic couplings studied below.
}

\section{Study of short-range epistatic model} 
We now consider stochastic evolution under a fitness model with  strong and spatially organized epistatic couplings,
\begin{equation}\label{fit6}
F(\textbf{s}) = f\sum_{i=0}^{L-1}s_i \,s_{i+1}\ ,
\end{equation}
where $s_{L}\equiv s_0$ (circular chromosome); $F$ is maximal for the two genomes $(-,-,...,-)$ and $(+,+,...,+)$. {As for the recombination correlation, we will choose either $c_{ij}=\frac 12(1-\rho^{|i-j|})$, which increases from 0 to $\frac 12$ with the distance between the loci, or $c_{ij}=\frac 12$ independently of the sites $i\ne j$. While the latter choice is less realistic, it does not affect the qualitative behavior of the model as we will see below, and allows for mathematical analysis in the $N\to\infty$ limit. Our goal is to understand how the distribution of alleles in the population change as a function of $\mu$ and $r$. Hereafter, we define $x=\mu/f$, $y=r/(2f)$.  }


\section{Paramagnetic phase} Due to the absence of local biases ($f_i=0$), the fitness is left unchanged by a global flip of all loci, $s_i\to-s_i$. In the absence of spontaneous symmetry breaking and in agreement with \mref{eq:QLE}, we assume that $\chi_i=0$ for all sites $i$. In this paramagnetic (PM) regime we look for a stationary and translation--invariant solution, $\chi_{ij}=\chi^{st} \big( d=|i-j|\big)$, to the coupled equations \mref{eq:corr} owing to the periodic boundary conditions in (\ref{fit6}). We find that, for all $d\ge 1$,  
\begin{eqnarray}
\chi^{st}(d+1) \!\!\!&=&\!\!\! \big[ 3  \chi^{st}(1)+a\big]\chi^{st}(d) - \chi^{st}(1)^2\, \delta_{d,1}  \\
&-&\frac 12\sum_{k=0}^{d-1}\sum_{s=0,1}\chi^{st}(k+s)\chi^{st}(d-k-1+s)   \nonumber \\
&-& \sum_{k=1}^{\infty} \sum_{s=0,1}\chi^{st}(k+s)\chi^{st}(d+k+1-s) \, , \nonumber
\label{eq:approx_epi_only}
\end{eqnarray}
where $a\equiv 2x+\frac y2$. To obtain a tractable approximation for $\chi^{st}(d)$ we further neglect all the terms with $k>d+1$ in the r.h.s. (last term) of (\ref{eq:approx_epi_only}). This approximation is based on the assumption that correlations decay quickly with the distance between loci, which will be checked a posteriori. Within this approximation the generating function $\tilde{\chi}^{st}(z)\equiv\displaystyle{\sum_{d\ge 0}\chi^{st}(d)z^d}$ is given by $G(z,u=0)$, where
\begin{equation}\label{chi780}
G(z,u) = \frac{z(3\chi^{st}(1)+a+4u) \pm \sqrt{\Delta(z,u)}}{1+z^2} \, ,  
\end{equation}
and $\Delta(z,u) = z^2 (3\chi^{st}(1)+a+4u)^2  +(1+z^2)[(1-u)^2-2z(1-u)(2\chi^{st}(1)+4u+a)-2z^2\chi^{st}(1)(\chi^{st}(1)+2u)]$; notice that $\chi^{st}(1)=d\tilde \chi^{st}(z)/dz|_{z=0}$.

\begin{figure}
	\centering
	\includegraphics[width=\linewidth]{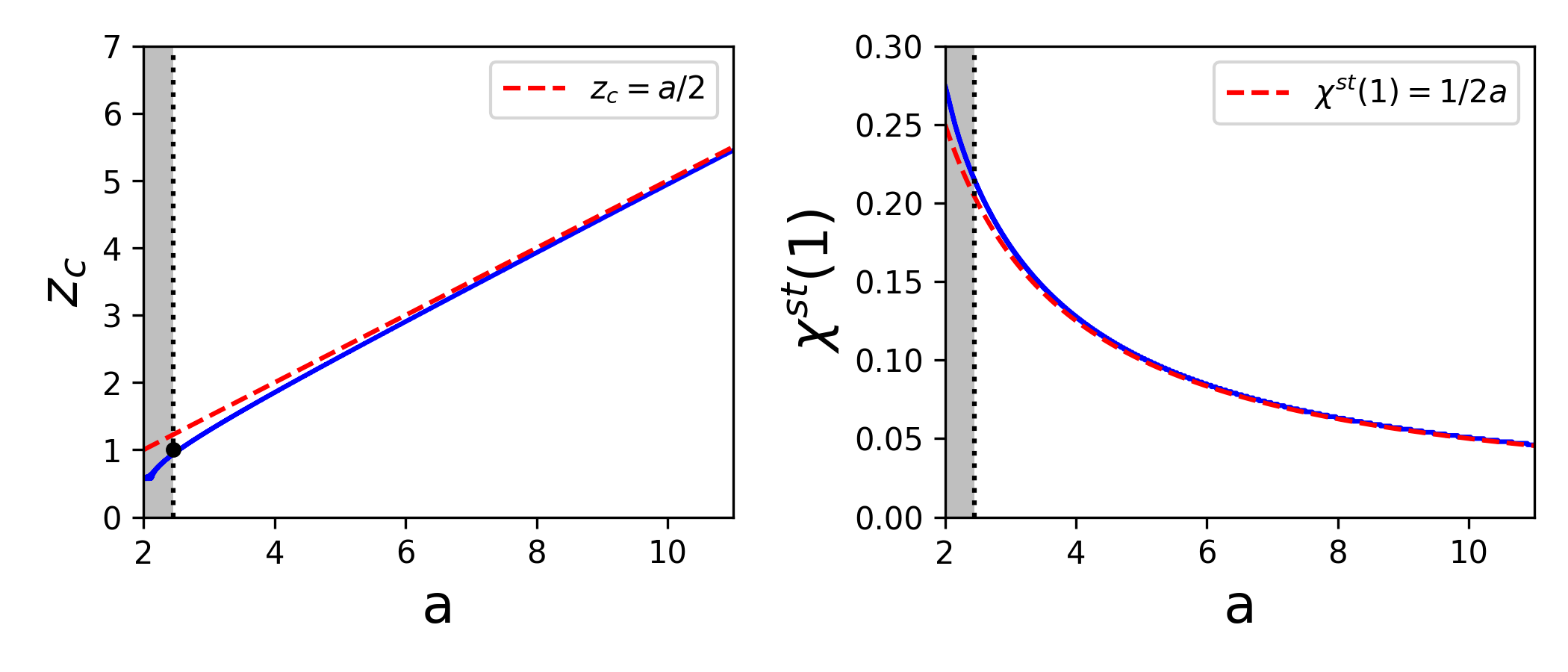}
	\caption{Radius of convergence $z_c$ (left) and corresponding value of $\chi^{st}(1)$ (right) vs. $a=2x+\frac y2$ in the PM regime.  Red dotted lines show the large--$a$ asymptotic behaviours.}
	\label{fig:kappa}
\end{figure}

The singularities of $\tilde{\chi}^{st}(z)$ control the asymptotic behaviour of $\chi^{st}(d)$ \cite{zbMATH05485323}. More precisely, if $\tilde{\chi}^{st}(z)$ behaves as  $(z_c-z)^\beta$ close to its singularity, then  
\begin{equation}\label{chi781}
\chi^{st}(d)\sim z_c^{-d}\, d^{-(\beta+1)}
\end{equation} 
for large $d$; here, the square root in \eqref{chi780} entails that $\beta=\frac 12$. To locate the singularity $z_c$ we identify the roots of $\Delta(z,0)$. As $\Delta$ is a polynomial of degree 4, it has two real roots and two conjugated complex roots, with modulus lower than 1. For values of $a>a_c\simeq 2.542$, there is one appropriate value for $\chi^{st}(1)$, which depends on $a$, such that the two complex roots merge on the real axis. This choice, accompanied by a change of sign in front of $\sqrt{\Delta}$ in (\ref{chi780}), removes the corresponding singularity and ensures that the radius of convergence $z_c$ of $G$ is larger than unity, as required for the exponential decay of $\chi^{st}(d)$. The radius of convergence $z_c$ and the corresponding value of $\chi^{st}(1)$ are shown as functions of $a$ in Fig.~\ref{fig:kappa}; for large $a$, $z_c\simeq \frac a2\gg 1$. Therefore the correlation function $\chi^{st}(d)$ \eqref{chi781} decays very quickly with $d$, as expected for large mutation or recombination rates.  For $a<a_c$ removal of the singularity corresponding to the small-modulus complex roots is not possible, indicating the presence of a qualitatively different regime we will study below.

\begin{figure}
	\centering
	\includegraphics[width = \linewidth]{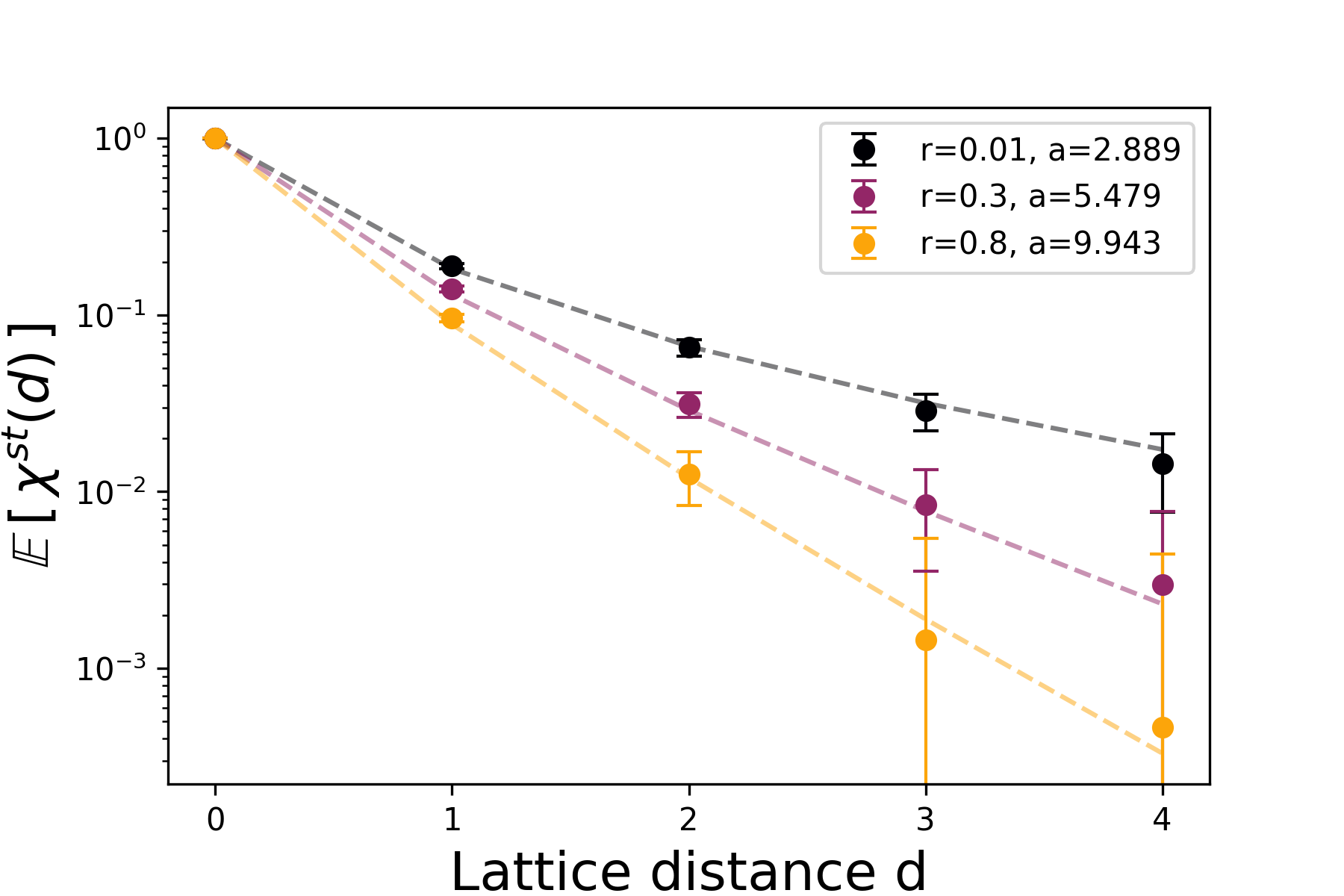}
	\caption{Average correlations in the PM phase for a finite population (dots) against the  value computed from Gaussian closure eqs. \mref{eq:QLE,eq:corr} (dashed lines) through Euler approximation, for three recombination rates $r$. Parameters: $f=0.028, L=50, \mu= 0.03, N=10^4$, $c_{ij}=\frac 1 2 (1-\rho^{|i-j|})$ with $\rho=0.6$; results are obtained by averaging over $5000$ generations.}
	\label{fig:corr_finite}
\end{figure}

We compare in Fig.~\ref{fig:corr_finite} the average values of the correlations $\chi_{ij}$ obtained by simulating a finite population with the ones computed above using the generating function $\tilde{\chi}^{st}(z)$. The agreement is excellent for small $d$, and more difficult to assess at large $d$ due to the finite-size and noise effects in the simulations.

\section{Critical line and spontaneous symmetry breaking}
\begin{figure}
		\centering
		\includegraphics[width=\linewidth]{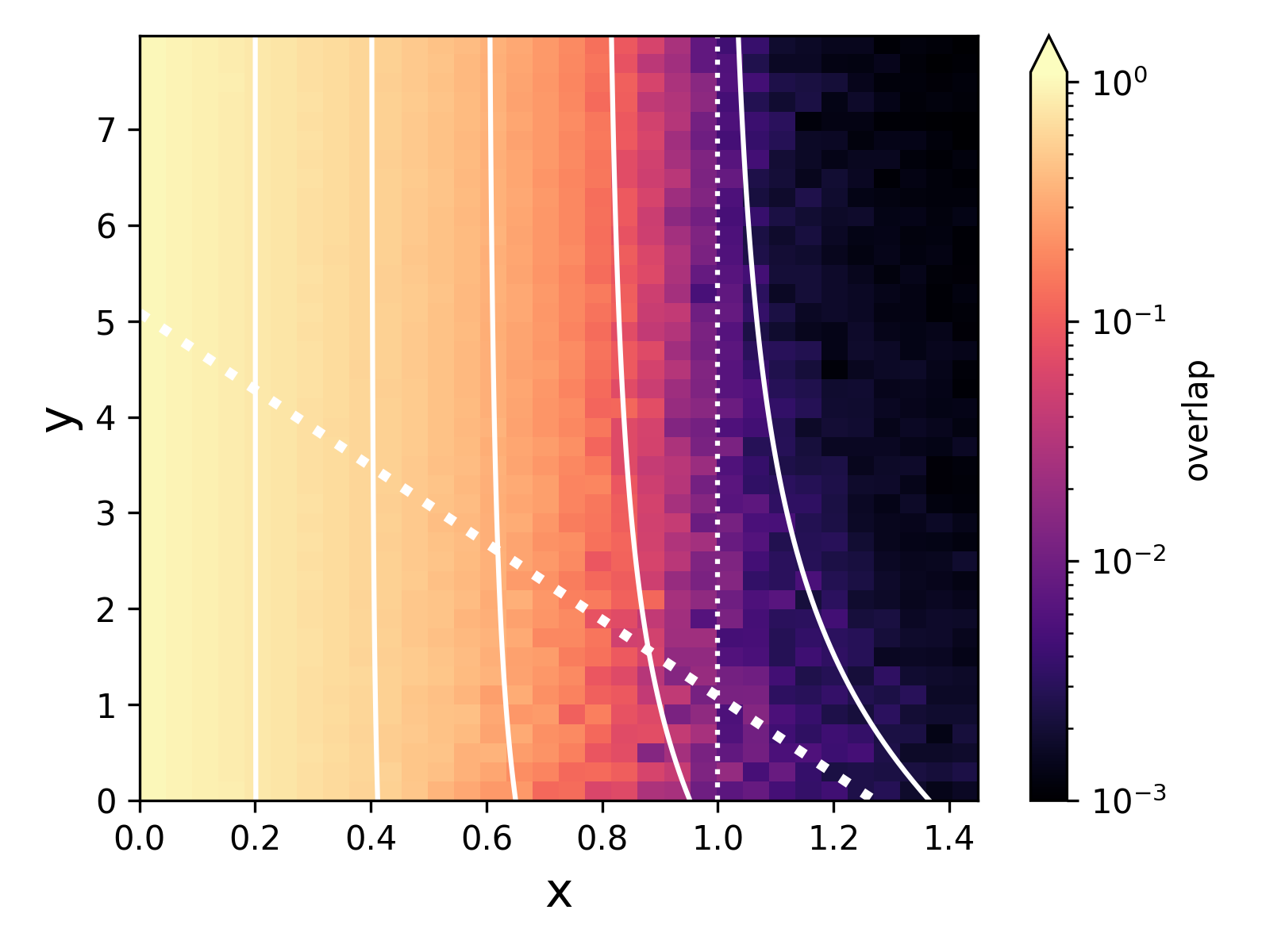}
		\caption{Phase diagram for genomes evolving under fitness (\ref{fit6}) in the $(x=\mu/f,y=r/(2f))$ plane. The heatmap shows the Edwards--Anderson overlap $q$, see text and color bar, for a simulated population of $N=10^4$ individuals in the case of non-homogeneous recombination matrix $c_{ij}=\frac 1 2 (1-\rho^{|i-j|})$ with $\rho=0.6$. The theoretical white lines were derived for recombination correlation $c_{ij}=\frac 12$, see main text. The critical line $x=2\tilde{\chi}^{st}(1)-2\chi^{st}(1)-1$  (rightmost solid curve) separates PM (right) and FM (left) phases, and approaches the $x=1$ line (thin dashed) when $y\gg 1$.
		Contour lines corresponding to $\chi_{FM}^2=0.2, .0.4, 0.6, 0.8$ (from right to left) are shown in the FM phase.
		The thick dashed line stands for $a=a_c\approx2.542$, showing that $a>a_c$ throughout the PM phase.} 
		\label{fig:phase_diagram}
	\end{figure}

The condition $a>a_c$ is necessary for the correlations to decay with distance, but it is not sufficient for the paramagnetic phase to hold. To investigate the onset of non-zero  $\chi_i$'s, we study the stability of the PM solution found above against small fluctuations of the frequencies $\chi_i$. To do so we linearize  eqn.~\eqref{eq:QLE} at fixed correlations $\chi_{ij}=\chi^{st}(|i-j|)$, and obtain $\dot{\chi}_i = -2f \sum_j {\cal M}_{ij}\chi_j$, where the entries of the $\cal M$ matrix read
	\begin{multline}
	{\cal M}_{ij} =\big[ 2\chi^{st}(1) +x\big] \delta_{ij}+\\- \frac 12 \bigg[\chi^{st}(|j-i+1|)+\chi^{st}(|j-i-1|)\bigg]\,,
	\end{multline}
	Being $\cal M$ translation invariant it can be diagonalized by Fourier plane waves, and its eigenvalues $\lambda_n$ are given by, for $n=0,1,...,L=1$,
		\begin{multline}
	\lambda_n = 2 \chi^{st}(1)+x+\\-\left[2\, \mathrm{Re} \, \tilde{\chi}^{st}\left(e^{-i\frac{2\pi}{L}n}\right)-1\right]\cos\left(\frac{2\pi n}{L}\right) \,, 
	\end{multline}
where $\tilde{\chi}$ is the generating function in \eqref{chi780}. The PM phase is stable as long as all the eigenvalues $\lambda_n$ are strictly positive. The vanishing of $\lambda_0$, the lowest eigenvalue of $\cal M$, thus defines the boundary of the paramagnetic phase. This critical line is shown in the $(x,y)$ plane in Fig.~\ref{fig:phase_diagram}, and corresponds to the condition $x = 2 \tilde{\chi}^{st}(1)-2\chi^{st}(1)-1$. We compared our analytical results, valid for $N\to\infty$, with simulations for a large but finite population size. We report the values of the Edward-Anderson overlap $q=\frac{1}{L}\sum_{i}\chi_i^2$ in Fig.~\ref{fig:phase_diagram}. We observe that $q$ takes very low values to the right of the critical line where the paramagnetic phase is predicted to be stable, which justifies a posteriori our approximation above. To the left of the critical line, $q>0$, which signals the onset of a ferromagnetic phase; the mutation rate is too weak to ensure that the average values of the $s_i$'s are zero.

\section{Ferromagnetic phase}

In the broken symmetry phase, hereafter referred to as ferromagnetic (FM), we expect the allele frequencies not to vanish any longer, and be equal to $\chi_i=\chi_{FM}$, up to a global reversal symmetry. Repeating the procedure  followed in the PM phase we again consider the generating function $\tilde \chi^{st}(z)$ for the correlations $\chi^{st}(d)$ at distances $d\ge 1$. According to eqn.~(\ref{eq:QLE}) the average allele frequency reads
\begin{eqnarray}
\label{eq:self_consistent}
\chi_{FM}^2 &=& 1-x + 2\sum _{d\ge 2} \chi^{st}(d) \nonumber \\
&=& x- \big[ 2\tilde{\chi}^{st}(1)-2\chi^{st}(1)-1\big] \ ,
\end{eqnarray}
where the generating function of the correlations is now given by $\tilde{\chi}^{st}(z)= G\big( z, \chi_{FM}^2\big)$, see (\ref{chi780}). As in the PM case, $\chi^{st}(1)$ is appropriately chosen to remove the spurious singularities inside the unit circle in the complex $z$--plane. By solving numerically the self-consistent eqns.~(\ref{chi780}) and (\ref{eq:self_consistent}) we were able to compute $\chi_{MF}^2$ throughout the ($x,y$) plane. As expected, $\chi_{MF}$ is non zero to the left of the border line of the region of stability of the PM phase; contour lines for $\chi_{FM}^2$ in the FM phase are shown in Fig.~\ref{fig:phase_diagram}. Notice that, for a large range of parameter values, correlations decays very fast with distance {as expected in the QLE phase}, and $\chi_{FM}\simeq\sqrt{1-x}$ according to eqn.~(\ref{eq:self_consistent}).{ In addition, we have numerically checked that condition (\ref{sc}) is met everywhere in the FM and PM phases, {\em i.e.} $2\sum _{d\ge 1 } \chi^{st}(d) <1$.}

\subsection{Short term dynamics: allele domains}
\begin{figure}
	\centering
	\includegraphics[width = \linewidth]{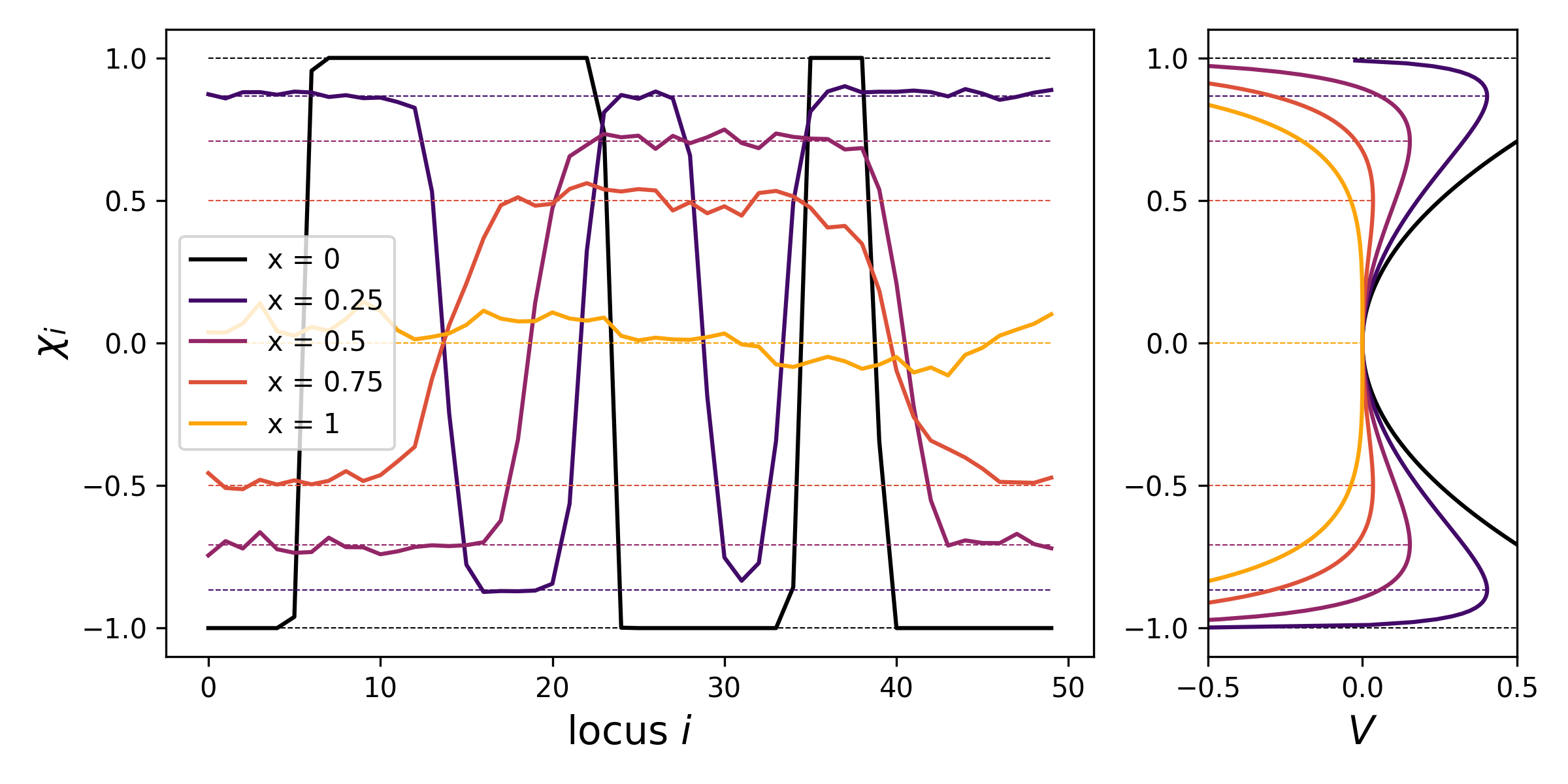}
	\caption{Left: Allele frequencies $\chi_i$ for one population evolved over 500 generations in the FM phase and for different values of the ratio $x$. Parameters: $f=0.028, N=10^4, L=50, r=0.85$ and $c_{ij}=\frac 1 2 (1-\rho^{|i-j|})$ with $\rho=0.6$. Dashed horizontal lines show $\pm\sqrt{1-x}$. Right: effective potential $V(\chi)$ for respective value of $x$ shown in the legend and horizontal lines locating the maxima of the potential in $\chi_{FM}=\pm\sqrt{1-x}$.}
	\label{fig:spin_lattice}
\end{figure}

Simulations show that, after a transient, the $\chi_i$'s reach apparently stationary values, with up and down domains of  height $\pm \chi_{FM}$, see Fig.~\ref{fig:spin_lattice}. Domains are of variable lengths, and are separated by characteristic domain walls.

The shape of walls separating contiguous domains can be analytically characterized in the $r \to \infty$ limit. For $x$ close to 1 and $L\gg 1$, a smooth, continuum limit for the index $i\to z$ can be taken. In this approximation, we  write $\chi_{i+1}+\chi_{i-1}-2\chi_i\simeq \chi''(z)$ in the stationary regime, and find, according to (\ref{eq:QLE}),
\begin{equation}
\chi''(z) = \frac{2\,x\,\chi(z)}{1-\chi^2(z)}-2\,\chi(z) \ ,
\label{eq:mech}
\end{equation}
After integration, we obtain $\frac{1}{2}\, \chi'(z)^2 +V\big(\chi(z)\big) = E$, where $E$ is a constant and the potential is $V(\chi) = x \ln\left(1-\chi^2\right)+\chi^2$.
Hence, finding the stationary profile of the spin average values is equivalent to a classical mechanics problem for a point-like particle of coordinate $\chi$ and unit mass, moving in the potential energy $V$ over fictitious time $z$. The potential $V(\chi)$ is symmetric in $0$, has two maxima in $\chi=\pm \chi_{FM}(x)$ and a local minimum in $\chi=0$ see Fig.~\ref{fig:spin_lattice}(right). 
The trajectories $\chi(z)$ with conserved energy $E=V\big(\chi_{FM}\big)$ define the domain wall profiles. These profiles depend on the value of $x$, and are in good agreement with Fig.~\ref{fig:spin_lattice}(left).



\subsection{Long-term dynamics: domain coarsening}
\begin{figure}
	\centering
	\includegraphics[width=\linewidth]{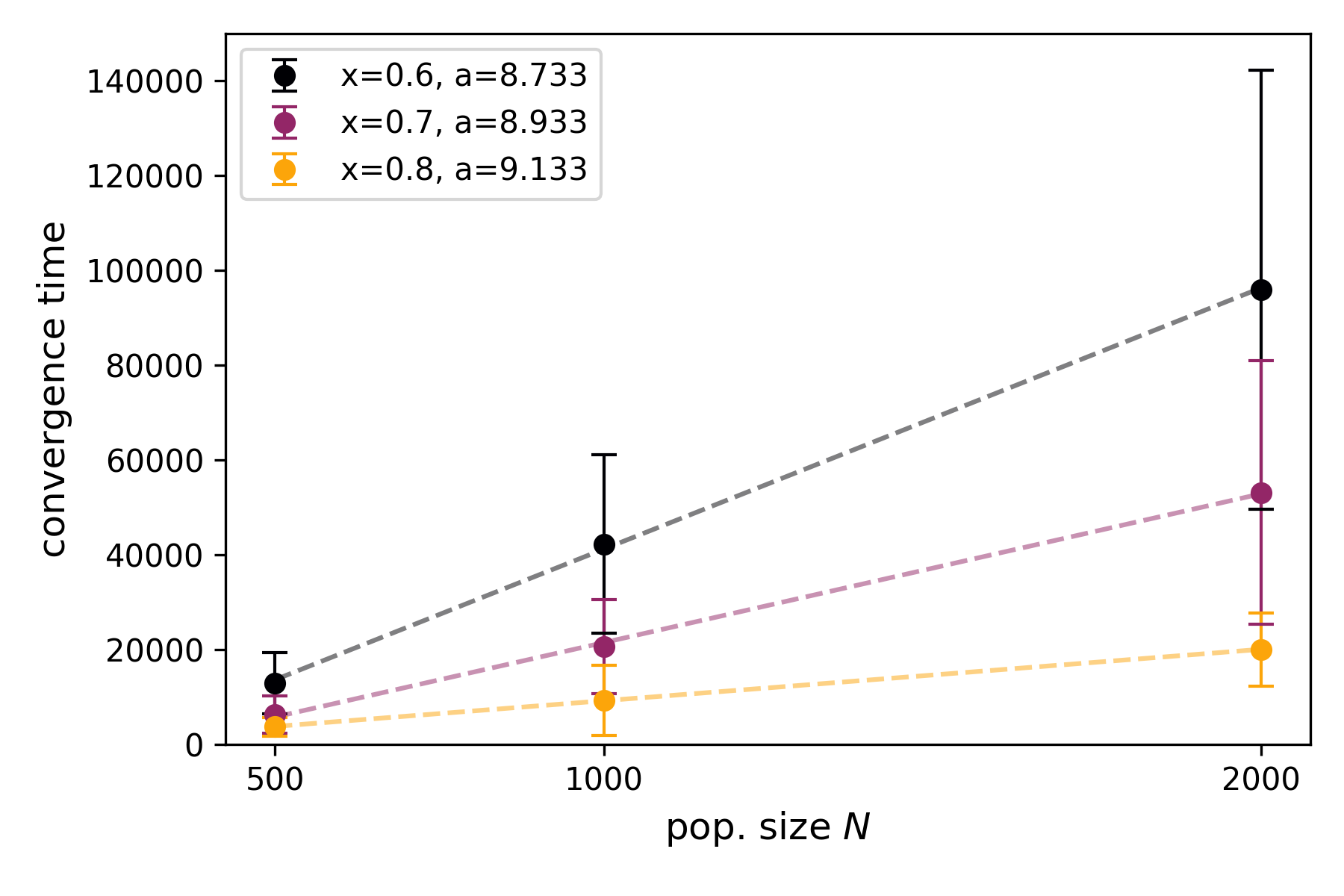}
	\caption{FM phase: Nb. of generations required for convergence towards the uniform frequency state $\chi=\chi_{FM}$ vs. population size $N$ for different values of $x$. The dashed lines stand for the linear regression of the data. Each point is the average over 10  simulations. Parameters: $r=0.85$, $f=0.028$, $L=50$, $c_{i\ne j}=\frac 12$. }
	\label{fig:coarsening}
\end{figure}

On longer time scales, the domain walls described in Fig.~\ref{fig:spin_lattice} undergo a diffusive process, due to the noise resulting from finite (population) size sampling. Domain walls merge when they encounter, and the system will eventually settle into a final state with uniform frequency $\pm \chi_{FM}$. The characteristic time the system needs to converge into the uniformly magnetized state is shown in Fig.~\ref{fig:coarsening}. It depends on the population size $N$ and on the genome length $L$, and is expected to scale as
\begin{equation}\label{sca}
\tau \sim \tau_0\times L^2\times N \ ,
\end{equation}
where $\tau_0$ depends on the intensive parameters, such as the mutation/fitness ratio $x$. The scaling in (\ref{sca}) stems from the following simple argument. The typical size of domains grow with time as $\sim t^{1/2}$ according to the Allen-Cahn theory of coarsening for non-conserved fields \cite{bray}. Hence a single domain will occupy the whole genome after a time given by $L^2/D$, where $D$ is the diffusion coefficient. Here, $D$ is expected to be inversely proportional to the population size $N$. This scaling is corroborated by the numerical results reported in Fig.~\ref{fig:coarsening}.

	\section{Conclusions}
	
	We have here introduced a Gaussian closure scheme for the dynamics of multi-allele correlations across an infinite-size population of genomes evolving under recombination, mutation and selection. Informally speaking, the distribution of genomes at time $t$ is described as a cloud of covariance matrix $\chi_{ij}(t)$ around an "average" genome (set of allele frequencies) $\chi_i(t)$. These 1- and 2-point correlations obey a set of $\frac 12 L(L+1)$ coupled deterministic, non-linear first order differential equations, where $L$ is the genome length. For the sake of simplicity, we have assumed that alleles can be in one out of two possible states. While this assumption may already lead to quantitative behaviors similar to real systems \cite{shekhar2013spin} it can be easily extended to multi-categorical (Potts) variables \cite{Gao_2019}.
	
	{Our scheme can be formally seen as a practical way to investigate the QLE regime beyond the large--$r$ perturbative approach initiated by Kimura \cite{Kimura65}. While our eqns.~\eqref{eq:corr} allow us to find back the QLE expressions for $\chi_{ij}$ at high recombination \cite{hongli}, they include contributions to all orders in powers of $1/r$ and can be exploited for generic  recombination and mutation rates, as well as for any fitness function, 
		and unveil non-trivial behaviors taking place at intermediate values of their defining parameters. An illustration is provided by the model analyzed in detail in this paper, in which two competing genomes, far away from each other in terms of allele composition have maximal and equal fitness. We find, at low enough mutation rate, a phase transition from a paramagnetic (PM) phase in which the  distribution is broad and encompasses the two fittest genomes to a ferromagnetic (FM) regime, in which one of the genomes eventually dominates. We stress that, both in the PM and FM phases, correlations decay exponentially with the distance between the alleles along the genome, which is also their distance on the epistatic interaction graph. Hence, both phases  formally belong to the QLE regime, even when $r$ is not large and $\mu$ is not small. However, contrary to a usual, informal definition of QLE, the genome distribution is not parametrized by the allele frequencies $\chi_i$ only, but also by the correlations $\chi_{ij}$, a consequence of the finiteness of $r$ and $\mu$.
	}

	Some aspects of the present work would deserve further investigation. {First, while determining the range of validity of closure schemes is generally a complicated issue, {\em e.g.} in the case of small chemical systems \cite{closurechem}, we have here derived some sufficient condition, see eqn. (\ref{sc}). 
		This condition will break down if the genome distribution is multi-modal and correlations are not small, as is the case in the Clonal Competition (CC) regime 
		\cite{Neher_clones}. However, even in CC, we may expect that our scheme remain adequate to describe locally the distribution of genomes attached to a clone. Therefore, it would be interesting to consider mixtures of Gaussian Ans\"atze, each describing a clone, an approximation which could be numerically tractable for sufficiently small numbers of clones and alleles. }
	
	Second, our approach offers, in principle, a practical way to infer the fitness determinants ($f_i, f_{ij}$) from the statistics of sequence data. Approaches to interpret correlations are needed as epistatis is notoriously difficult to infer from observation of mean fitness alone \cite{McCandlish}. Recently, the present Gaussian scheme was successfully used to reconstruct epistatic contributions from synthetic data \cite{hongli}. We hope the present approach will help put the inverse statistical approaches to protein sequence data \cite{coccorev} onto firm population genetics grounds.

	\acknowledgments
	We are grateful to E. Aurell, V. Dichio, G. Semerjian, H-L. Zeng for very useful discussions. E.M. is funded by an ICFP fellowship of the Physics Department at Ecole Normale Sup\'erieure. We acknowledge financial support from the ANR-17-CE30-0021 RBMPro grant.
{
\appendix
\section{Derivation of dynamical equations for the first and second moments}

Under the selection, mutation and recombination steps described in the main text the distribution of genomes at time $t$, $P({\bf s},t)$, obeys the Fokker-Planck-like equation:
\begin{eqnarray}
\dot P({\bf s},t) &=& \big( F({\bf s}) - \langle F\rangle\big) \, P({\bf s},t) + \mu \sum_{i=1}^L \big( P({\bf s}^{(i)},t) -P({\bf s},t) \big) \nonumber \\
+ & r &\sum _{\boldsymbol\xi,{\bf s}'} C(\boldsymbol\xi) \big(P({\bf s}^m,t)P({\bf s}^f,t)-P({\bf s},t)P({\bf s}',t)\big)  ,
\end{eqnarray}
which is identical to eqn (7) in \cite{Neher_QLE}. In the above equation, ${\bf s}^{(i)}$ denotes genome $\bf s$ in which allele $i$ has been flipped, $s_i\to-s_i$. By multiplying the above equation by $s_i$ or $s_is_j$ and summing over all genomes $\bf s$ we obtain the evolution equations for the moments $\chi_i$ or $\chi_{ij}$, that is, $
\dot \chi_i = \langle s_i \big(F({\bf s}) - \langle F\rangle \big)\rangle- 2\mu \chi_i$ and 
$\dot \chi_{ij} = \langle (s_i-\chi_i)(s_j-\chi_j) \big(F({\bf s}) - \langle F\rangle \big)\rangle- (4\mu+r c_{ij}) \chi_{ij}$,
where the fitness function is defined in (\ref{fit1}). The above equations coincide with eqns (8) and (10) in \cite{Neher_QLE}. We are left with the evaluation of the averages $\langle\cdot \rangle$ in the above equations. After replacement of quadratic terms $s_k^2$ with 1, the polynomial to be averaged is of degree zero or one in each variable $a_k=s_k-\chi_k$. We then forget about the discrete nature of the $a_k$'s, and take them as multivariate Gaussian variables with zero means and covariance matrix $\chi_{k\ell}$. Hence all odd moments vanish, and even moments can be computed from Wick's theorem, e.g. $\langle a_k a_\ell a_m a_n\rangle=\chi_{k\ell}\chi_{mn}+ \text{2\ other pairings}$. A systematic evaluation of all terms lead to (\ref{eq:QLE}) and (\ref{eq:corr}).

\section{Contributions to $k^{th}$ moment} The approximation scheme describe in Appendix A can be used to compute moments of any order, such as $\chi_{i_1,i_2,...,i_k}=\langle\prod_{\alpha=1}^k (s_{i_\alpha}-\chi_{i_\alpha})\rangle $ with $k\ge 3$. In the stationary regime, we obtain
\begin{eqnarray}
\chi_{i_1,i_2,...,i_k} &=&\frac 1{G_k} \bigg\langle \bigg\{\prod_{\alpha=1}^k a_{i_\alpha} - \sum_{\alpha=1}^k (a_{i_\alpha}+\chi_{i_\alpha})\langle\prod_{\beta(\ne \alpha)}a_{i_\beta}\rangle\bigg\}\nonumber \\
&\times&  \big(\sum_ i \hat f_i a_i + \sum_{i<j} f_{ij}(a_i a_j -\chi_{ij})\bigg\rangle 
\end{eqnarray}
where $\hat f_i + \sum _{j (\ne i)} f_{ji} \chi_j$, $G_k=2k \mu +r c_{i_1,i_2,...,i_k}$,  and $c_{i_1,i_2,...,i_k}$ is the probability (computed with $C(\boldsymbol\xi)$) that all $k$ alleles do not come from the same parent (father or mother). Careful inspection of the polynomial in the $a_k$ variables (after replacements $a_i^2 \to 1-\chi_i^2-2\chi_i a_i$ to enforce $s_i^2=1$) and of the effects of Wick's contraction shows that terms appearing in the average $\langle\cdot\rangle$ above are of the form $M\times C\times (\chi_{ij})^E\times I$, where $\chi_{ij}$ stands for a generic 2-point moment, and (1) the exponent $E$ is equal to $\frac \ell 2$, where $\ell$ is an integer comprised between $k-2$ and $k+2$, depending on the parity of $k$ and the term considered; (2) the combinatorial factor $C$ is equal to $(E-1)!!$; (3) the multiplicity $M$ is bounded by $L^3$; (4) the interaction term $I$ is a polynomial of degree $\le 2$ in two of the $\chi_{i_\alpha}$'s ($\alpha=1...k$), and is proportional to one local fitness term $\hat f_i$ or epistatic coupling $f_{ij}$.
}


\bibliography{ref3.bib}

\end{document}